\documentclass[11pt,twoside]{article}
\usepackage{asp2004}
\usepackage{psfig}
\usepackage{epsf}
\usepackage{graphics}
\usepackage{lscape}
\usepackage{aas_macros}
\usepackage{floatfig,epsfig}
\begin{document}
\initfloatingfigs

\title{AM CVn stars}
\author{Gijs Nelemans}
\affil{Institute of Astronomy, University of Cambridge, UK\\
and\\
Department of Astrophysics, Radboud University Nijmegen,  NL}

\begin{abstract}
  I review our observational and theoretical knowledge of AM CVn
  stars, focusing on recent developments. These include newly
  discovered systems, the possibility that two recently discovered
  extremely short period objects are AM CVn stars and an update on
  X-ray, UV an optical studies. Theoretical advances include the study
  of the details of both the donor and accretor, and the physics of
  the helium accretion discs. I review our (limited) knowledge of
  the formation of AM CVn stars and the apparent success of the now
  more than 25 year old suggestion that in these objects the mass
  transfer is driven by gravitational wave radiation losses. The
  exciting prospect of directly detecting these gravitational waves
  and the possibilities this brings conclude this contribution.
\end{abstract}
\thispagestyle{plain}

\section{Introduction}

AM CVn stars are binary systems that have very short orbital periods
(less than about one hour) and helium dominated spectra. The
prototype, with an orbital period of 17 min.  was discovered in 1967
\citep{sma67}. \cite{pac67} quickly proposed that this object was a very
short period binary with a degenerate, helium-rich donor and that the
mass transfer is driven by angular momentum loss due to gravitational
wave radiation (GWR). Because GWR depends very strongly on orbital
period, this interpretation suggests a rapidly dropping mass-transfer
rate as function of orbital period. Since 1967 another ten objects
have been discovered with periods up to 65 minutes. Four of them were
found in the last few years, as well as two extremely short period
objects that are possibly also AM CVn stars.  In this review I'll
first give a description of the properties of AM CVn stars, followed
by a short overview of our understanding of their formation and
evolution, recent observational developments, and highlight some
interesting questions and their possible answers.  Finally I discuss
AM CVn stars as GWR sources. For earlier reviews see
\citet{war95,sol95,sol03}.

\subsection{Fundamental properties of AM CVn stars}

\begin{table}[!ht]
\caption{Overview of observational properties of AM CVn stars}
\label{tab:overview}
\smallskip
\begin{center}
\hspace*{-0.5cm}
{\small
\begin{tabular}{|l|ll|l|l|l|l|c|c|}\hline
Name  & $P_{\rm orb}^a$ & & $P_{\rm sh}^a$ & Spectrum & Phot. var$^b$ & dist & X-ray$^c$ &
UV$^d$  \\ 
 & (s) & & (s) & & &  (pc) & &  \\
\hline \hline
ES Cet & 621  &(p/s) &  & Em$^{1,2}$ & orb &  & C$^3$X &  GI   \\
AM CVn & 1029 &(s/p) & 1051 & Abs & orb & 235$^4$ & RX & HI  \\
HP Lib & 1103 &(p) & 1119 & Abs & orb &  & X & HI \\
CR Boo & 1471 &(p) & 1487 & Abs/Em? & OB/orb & & ARX & I \\
KL Dra & 1500 &(p) & 1530 & Abs/Em? & OB/orb &  &  &  \\
V803 Cen & 1612 &(p) & 1618 & Abs/Em? & OB/orb &  & Rx & FHI   \\
CP Eri & 1701 &(p) & 1716 & Abs/Em & OB/orb & &  & H  \\
2003aw & ? & & 2042 & Em/Abs? & OB/orb &  & &    \\ 
SDSSJ1240-01 & 2242$^5$ &(s) & & Em & n &  & &    \\
GP Com & 2794 &(s) & & Em & n & 70$^6$ & ARX$^7$ & H$^8$I \\
CE315  & 3906 &(s) &  & Em & n & 77$^9$ & R(?)$^e$X & H$^{10}$  \\ 
 & & & & & & & & \\
Candidates & & & & & & & & \\\hline\hline
RXJ0806+15 & 321 &(X/p) & & He/H?$^{11}$ & ``orb'' &  &
$^{12-15}$CRX &  \\
V407 Vul & 569 &(X/p) &  & K-star$^{16}$ & ``orb'' &  & 
$^{17-19}$ARCRxX &  \\ \hline
\end{tabular}
}
\end{center}
{\small
$a$ orb = orbital, sh = superhump, periods from \citet{ww03}, see
references therein, (p)/(s)/(X) for photometric, spectroscopic, X-ray period.\\
$b$ orb = orbital, OB = outburst\\
$c$  A = ASCA, C = Chandra, R = ROSAT, Rx = RXTE, X = XMM-Newton \citep[see][for a
short review and references]{kns+04}\\
$d$  F = FUSE, G = GALEX, H = HST, I = IUE \citep[see][for
references]{sol95,war95}\\
$e$ 1RXS J131246.8-232118\\
References: 1 \citet{ww02}, 2  Steeghs et al, in prep., 3
\citet{str04c}, 4 C. Dahn, private communication, 5 Roelofs et al. in
prep. 6 \citet{tho03}, 7 
\citet{str04b}, 8 \citet{mms+03}, 9 Thorstensen, this conference, 10
\citet{gsd+03}, 11 \citet{ihc+02}, 12 \citet{ics+03}, 13
\citet{str03b}, 14 \citet{ihc+02}, 15 \citet{rhc02}, 16 Steeghs et
al. in prep., 17 \citet{chm+98},
18 \citet{str03}, 19 \citet{str04a}
}
\end{table}

The known AM CVn stars and two candidate systems are listed in
Table~\ref{tab:overview}. The table gives a very short overview of the
observational properties of the systems, with recent references (for
earlier work, see the previous reviews). As function of increasing
orbital period (and thus also as function of time as they evolve to
longer periods) they go through three distinct phases: (i) A
high-state phase (AM CVn and HP Lib, with $P \la$~20~min), with
low-amplitude photometric variations on various periods, among which
the orbital period and a slightly longer one, called the superhump
period \citep[see][]{war95}. The latter is the result of the fact that
the accretion disc becomes eccentric and starts to precess due to the
extreme mass ratio \citep{whi88}.  For AM CVn, the absorption lines
show variations on the beat frequency between orbital and superhump
frequencies. (ii) An outbursting phase, in which large optical
variability (up to 4 magnitudes) is seen (CR Boo, KL Dra, V803 Cen, CP
Eri, 2003aw with periods 20~$\la P \la$~40~min). During the bright
phases these systems resemble the the high-state absorption line
systems, whereas emission lines are visible during the quiescent
phases \citep[as far as we have seen these spectra, e.g.][]{gns+01}..
These systems are thought to have unstable discs, in analogy with the
hydrogen-rich dwarf novae \citep{to97}.
(iii) The longest period systems (SDSSJ1240-01, GP Com and CE315, with
$P \ga$~40~min) show no optical photometric continuum variability, but
their orbital periods are determined spectroscopically.  SDSSJ1240-01
also shows the He absorption lines of the accreting white dwarf
\citep{rgs+04}, while the continuum emission of GP Com and CE315 is
also shown to come from the white dwarf (Bildsten et al., in prep). In
these cases, however, the effective temperature of the accreting white
dwarf is too low to show He absorption. The low-state systems probably
have stable cool accretion discs.

\begin{figure}[t]
\begin{center}
\psfig{figure=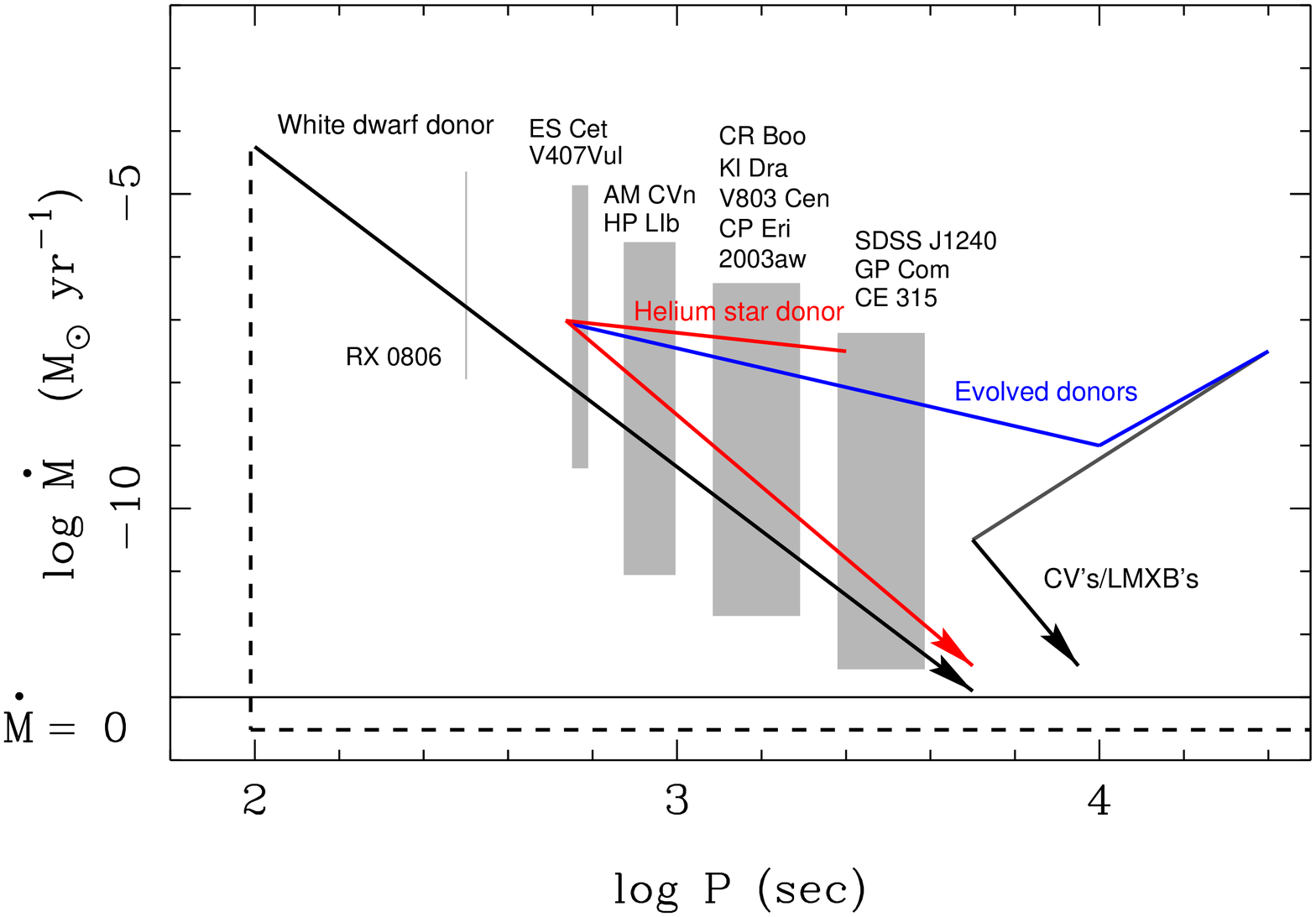,width=0.8\textwidth}
\end{center}
\vspace*{-0.5cm}
\caption{Formation paths of AM CVn stars (see text). The known systems
  (including the two candidates) are shown at their orbital period.}
\label{fig:P_Mdot}
\vspace*{-0.3cm}
\end{figure}

\subsection{Formation and evolution of AM CVn stars}

For a detailed discussion of the formation of AM~CVn systems I refer
to \citet{npv+00}, and references therein. There are three routes for
the formation of AM~CVn systems (see Fig.~\ref{fig:P_Mdot}): (i) via a
phase in which a double white dwarf loses angular momentum due to
gravitational wave radiation and evolves to shorter and shorter
periods to start mass transfer at periods of a few minutes after which
it evolves to longer periods with ever dropping mass-transfer rate
\citep[e.g.][]{pac67}. In order for the mass transfer to be stable and
below the Eddington limit, the mass of the donor must be small
\citep[e.g.][]{npv+00}. (ii) Via a phase in which a low-mass,
non-degenerate helium star transfers matter to a white dwarf accretor
evolving through a period minimum of about ten minutes, when the helium
star becomes semi-degenerate. After this minimum, the periods increase
again with strongly decreasing mass-transfer rate \citep{it91}.  (iii)
From cataclysmic variables with evolved secondaries \citep{phr01},
which, after mass loss of the evolved star has uncovered the He-rich
core, evolve rather similar to the helium star tracks.  These
formation paths are depicted in Fig.~\ref{fig:P_Mdot}, together with
the observed systems at their orbital periods. It is clear that in
order to distinguish between the different evolutionary scenarios,
more information than the orbital period is needed and that for the
currently observed systems all formation scenarios in principle are
viable (except for RXJ0806+15).

\subsection{Direct impact}

\addtocounter{figure}{1}
\begin{floatingfigure}{6cm}
\mbox{\includegraphics[angle=-90,scale=0.5,clip]{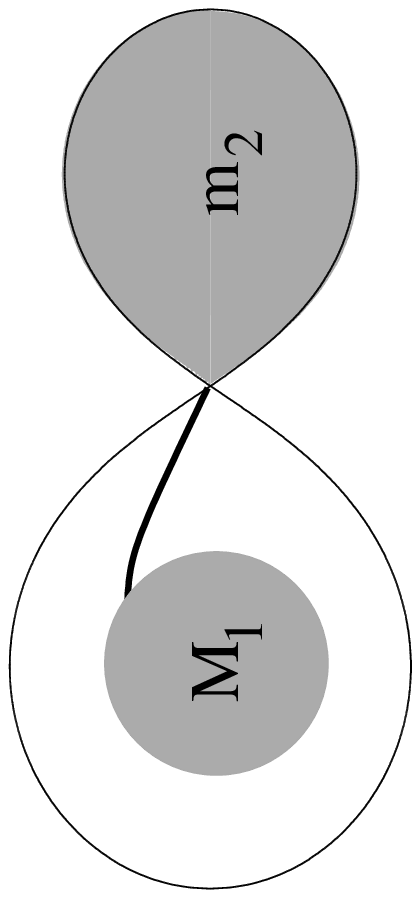}}\\
\vspace*{0.2cm}

\noindent{\fontsize{10}{11.5}\selectfont Figure 2. Schematic picture of the direct impact phase of accretion
  between two white dwarfs.}
\vspace*{0.3cm}
\label{fig:DI}
\end{floatingfigure}
\par
\noindent A special situation occurs at the onset of mass transfer between two
white dwarfs. The two objects are so close together that there is no
room for an accretion disc in the system and the stream of gas that
falls from the inner Lagrangian point towards the accreting white
dwarf impacts directly onto its surface \citep[][see Fig.
2]{web84,npv+00}. This has has important consequences for
the stability of the mass transfer and the AM CVn stars
formed from detached double white dwarfs \citep{npv+00,mns02}.

\vspace*{4.2ex}

\section{Recent observational developments}

\subsection{Low state systems}

\begin{figure}[hbt]
\begin{center}
\psfig{figure=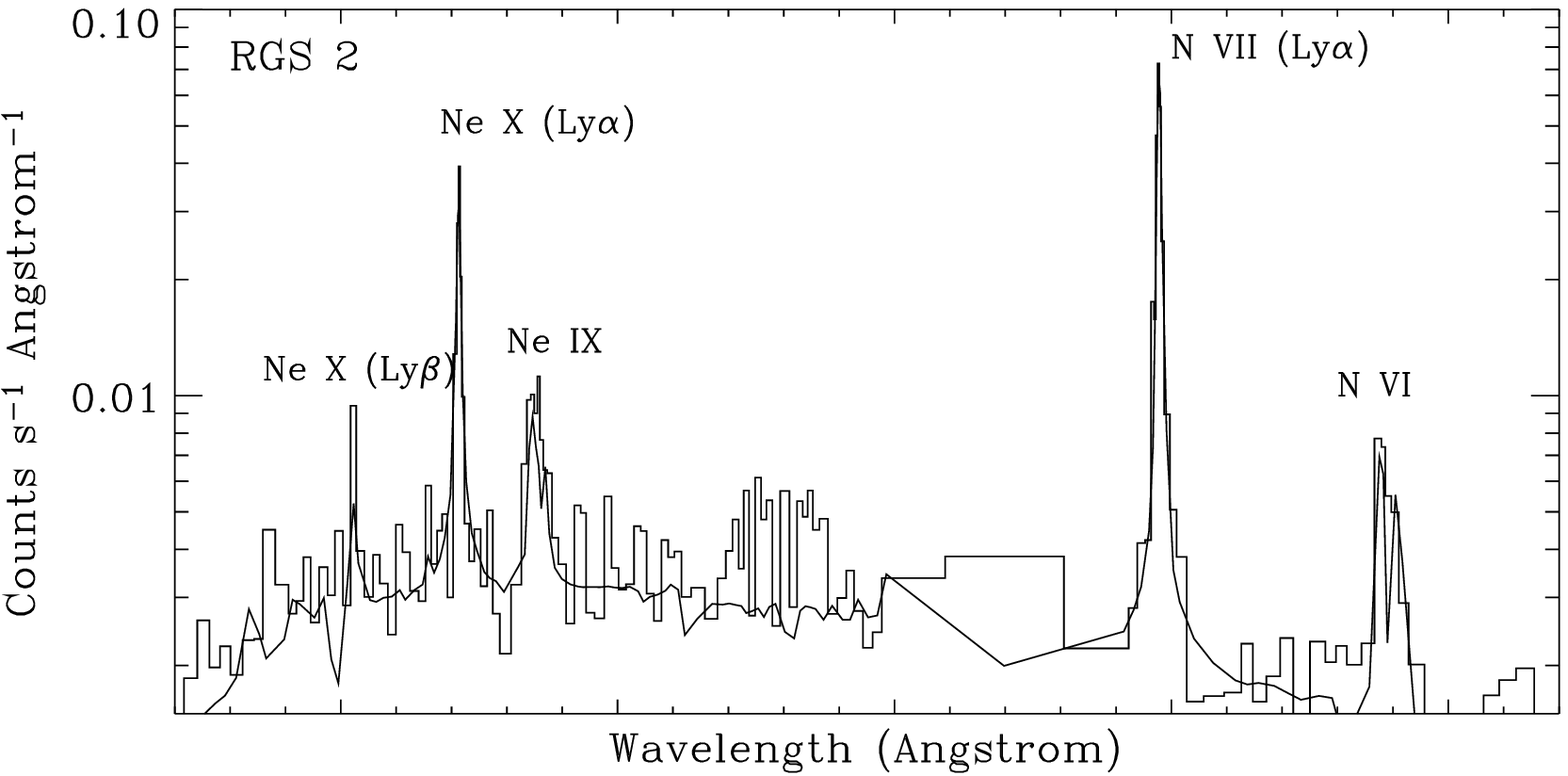,width=0.8\textwidth,clip}
\end{center}
\vspace*{-0.5cm}
\caption{XMM/Newton spectrum of GP Com, showing clear detection of N
  and Ne lines, confirming the peculiar chemical abundances.  From
  \citet{str04b}}
\label{fig:GP_Com_XMM}
\vspace*{-0.3cm}
\end{figure}

The low state systems are dominated by strong HeI emission lines, with
a weak continuum. In 2001 a new system was discovered, CE315
\citep[V396 Hya,][]{rrg+01}. It is virtually identical to GP Com in
terms of its optical spectrum.  Recent work on GP Com shows that the
mysterious central spike seen in the middle of the double peaked
emission lines indeed originates on the accreting white dwarf
\citep[][Steeghs et al., in prep]{mms+03}. This yields a radial
velocity amplitude of the accretor of GP Com of $11.7 \pm 0.3$ km/s
and for CE315 of $5.8 \pm 0.3$ km/s. Together with the hot spot
velocity this yield $q \approx 0.018$ for GP Com and $q \approx
0.0125$ for CE315.  The peculiar chemical abundance of GP Com, with
low abundances of heavy metals suggesting a low-metallicity origin and
highly over-abundant N (even compared to CNO processed material)
possibly accreted from an AGB companion \citep{mhr91} is confirmed by
further UV and X-ray studies \citep[][see
Fig.~\ref{fig:GP_Com_XMM}]{mms+03,str04b}.

A new system was found in the first data release of the Sloan Digital
Sky Survey: SDSSJ1240-01 \citep{rgs+04}. It shows the strong HeI
emission lines, but the continuum shows the DB spectrum of the
accreting white dwarf.  Follow-up spectroscopy yields an orbital
period of 2242 s (Roelofs et al., in prep.). No outburst have (yet)
been seen. A completely new feature is the presence of a quite strong
line at 5169\AA, probably from Fe.

\subsection{Outbursting systems}

The outbursting systems are viewed as being in the mass-transfer
regime where the He accretion disc is unstable \citep{to97}.  In their
bright phase they should behave similar to the high state systems and
in quiescence similar to the low state systems. Indeed in their bright
phase they look very much like the high state systems, but very few
quiescent spectra are available. The recent quiescent spectrum of CP
Eri \citep{gns+01}, confirms the similarity with the spectra of the
low-state systems, however with distinct differences.  There is no
sign of a central spike and the presence of Si lines suggest much
higher metallicity than GP Com and CE315. Two new objects are found
recently, interestingly, both in supernova searches. The first
\citep[SN 1998di, now called KL Dra,][]{jgc+98} showed a high state
spectrum in outburst and shows a photometric period of 1530 s., which
is interpreted as a superhump period suggesting an orbital period of
about 1500 s \citep{wcg+02}.  The other object, SN 2003aw,
\citep{cf03} interestingly in the discovery spectrum showed emission
lines, but seems to be discovered in a state of intermediate
brightness, as in May 2004 it reached $V \sim15$, about 1.5 mags
brighter than at its discovery (Woudt, private communication).
Follow-up photometry resulted in the discovery of a 2042 s.  period
\citep{ww03a}, again interpreted as a superhump period. VLT spectra of
this object show a spectrum that is very similar to that of SDSSJ1240-01 (Roelofs et al., in prep).  Recent papers on photometry of
outbursting superhump systems are \citet{pks+97,pwk+00,knb+00,ksm+04}
  
\subsection{High state systems}

The two high state systems, AM CVn and HP Lib are characterised by
broad, shallow helium absorption lines.  Relatively recently the
orbital period of AM CVn was confirmed spectroscopically
\citep{nsp00}. A similar study for HP Lib has not (yet) yielded
results. Extensive photometric campaigns of AM CVn and HP Lib have
shown a complex system of periodicities \citep{spk+99,pfr+02,sar+00}.
For a detailed discussion of the many types of variability in AM CVn
stars (and CVs in general) see \citet{war04} and Warner \& Woudt,
these proceedings.

\subsection{ES Cet}

\begin{figure}[t]
\begin{center}
\psfig{figure=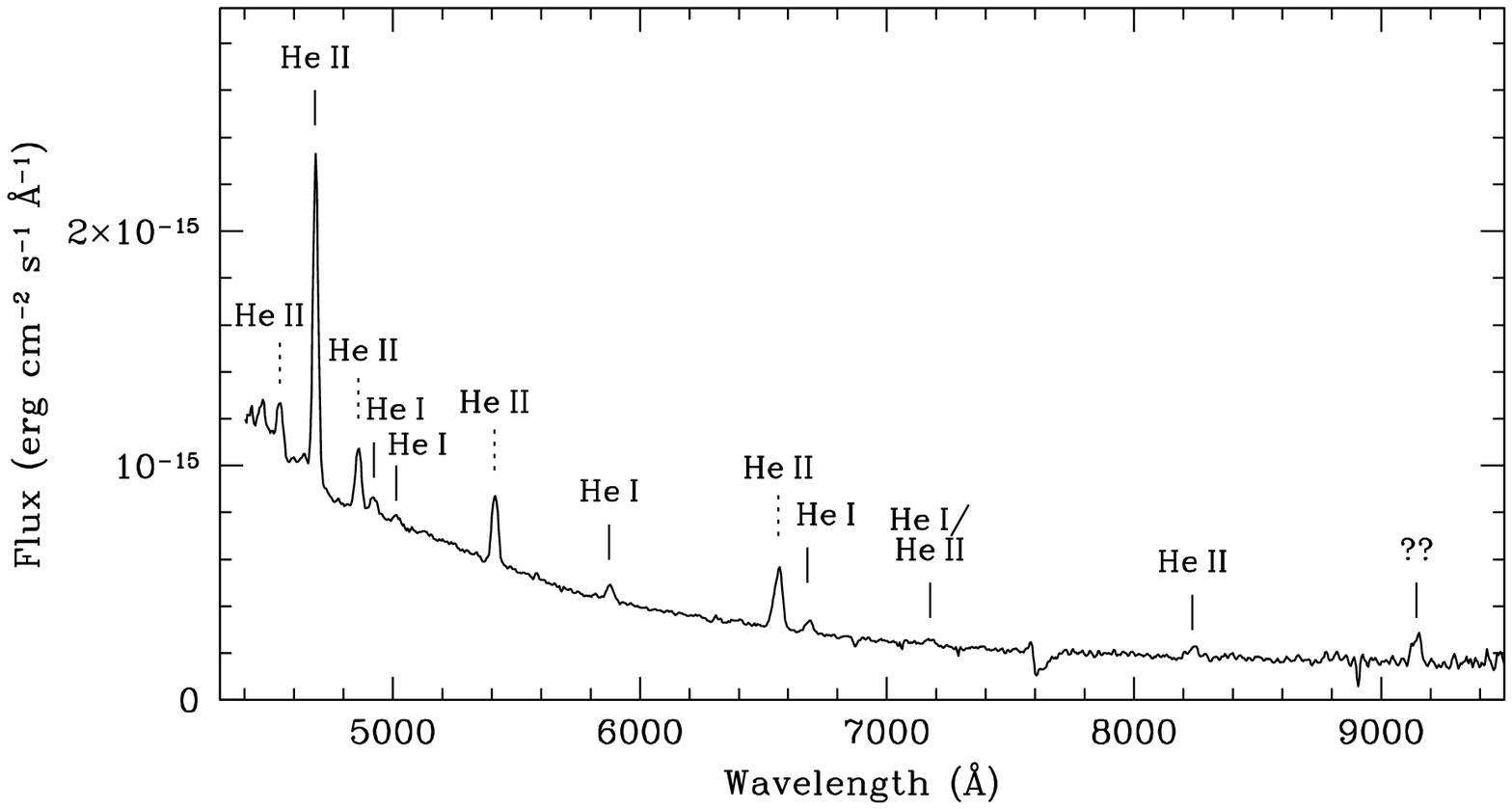,width=0.8\textwidth}
\end{center}
\vspace*{-0.5cm}
\caption{Average spectrum of ES Cet. From \citet{ww02}}
\label{fig:ES_Cet}
\vspace*{-0.3cm}
\end{figure}

ES Cet (KUV 01584-0939) was discovered as a strong emission line
system in the KISO survey, but in 2002 Brian Warner realized the ``H''
lines present in the spectrum of this peculiar CV were lines of the
HeII Pickering series (see Fig.~\ref{fig:ES_Cet}).  \citet{ww02}
discovered a 620 s.  period in their photometry, which is interpreted
as the orbital period, making ES Cet the shortest period binary known
to day.  Phase resolved spectroscopy clearly reveals the period in the
changing line profiles (Steeghs et al. in prep).  Its period is on the
edge of the period range where direct impact accretion is expected
\citep[see Fig.~4 of][]{nyp03}.

\subsection{Ultra-short period candidates}

\begin{figure}[t]
\begin{center}
\psfig{figure=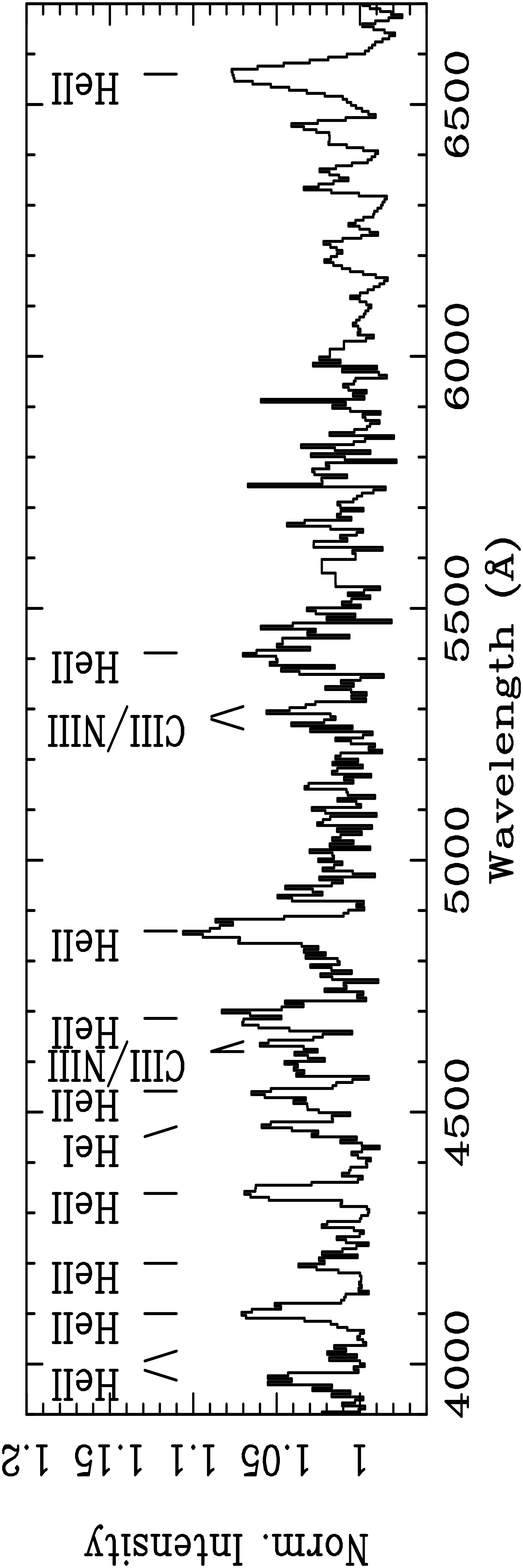,angle=-90,width=\textwidth}
\end{center}
\vspace*{-0.5cm}
\caption{VLT spectrum of RXJ0806+15. From \citet{ihc+02}}
\label{fig:RXJ0806}
\vspace*{-0.3cm}
\end{figure}

In the last few years a lot of attention has focused on two candidate
AM CVn stars: V407 Vul \citep[RXJ1914.4$+$2456][]{chm+98} and
RXJ0806.3$+$1527 \citep{ihc+02,rhc02} which were discovered as
\textit{ROSAT} X-ray sources \citep{mhg+96,ipc+99}. They show an
on-off X-ray light curve with periods of 9.5 and 5.4 minutes and very
soft X-ray spectra.  These periods are also found in the optical, but
out of phase \citep{rcw+00,ics+03}.  No other periods are found in
their light curves. The optical spectra are peculiar.  \citet{ihc+02}
showed a spectrum of RXJ0806+15 with very weak emission lines, which
they interpreted as HeII lines (see Fig.~\ref{fig:RXJ0806}). The fact
that the even lines of the series (which are very close to the Balmer
lines) are stronger than the odd lines can also be interpreted as
evidence for the presence of some H.  In the optical and IR the
spectral energy distribution is consistent with a single hot black body
\citep{rbs04}.  The optical spectrum of V407 Vul shows the spectrum of
a late G-star (Steeghs et al., in prep.). If not the same object the
G-star is very close to the object (within 0.03", Marsh et al., these
proceedings), the lack of any detectable radial velocity and emission
lines appears at odds with the observed X-ray emission.  \citet{isr04}
have reported detection of linear polarisation from RXJ0806+15. A very
interesting development is that for both objects it has become clear
that the periods are getting shorter \citep[][Israel et al. in
prep]{str03,hrw+03,str04a}, in contradiction with the expectation for
the secular evolution of AM CVn stars.

\subsection{Distances and Surveys}

Distances to AM CVn stars have been unknown for long. This is
currently changing (rapidly). John Thorstensen has determined 2: 70 pc
for GP Com \citep{tho03} and 77 pc for CE315 (Thorstensen, this
proceedings). The USNO parallax team (C. Dahn, private communication)
has measured the distance to AM CVn to be 235 pc. Distances to AM CVn,
HP Lib, CR Boo, V803 Cen and GP Com will be determined (confirmed) by
HST FGS parallaxes (Groot et al., in prep).

Another promising area of development for AM CVn stars are surveys.
Both general purpose surveys (in particular the Sloan survey) and
dedicated surveys to find more AM CVn stars. As a pilot project, a
survey was done of $\sim$40 square degrees at Galactic latitude
$\sim$15 degrees with imaging in two broad band filters (B and R) and
two narrow band filters (one on H$\alpha$ and one on the 6678 \AA\ He
emission line). The already known emission line systems (GP Com and
CE315) were clearly detected. However, among the few candidates no new
AM CVns were discovered (Groot et al., in prep). A larger, lower
Galactic latitude survey, including variability information will be
performed with the OmegaCam wide field imager on the VLT Survey
Telescope, the OmegaWhite survey (PI Paul Groot). The RApid Time
Survey looking for short-period photometric variables with the WFC on
the INT (Ramsay \& Hakala, in prep.) is another promising survey.

\section{Questions and partial answers}

AM CVn stars are interesting objects for a number of reasons, mainly
related to their short periods and the fact that the donor stars in
these systems have peculiar chemical composition. This makes it
possible to study some astrophysical processes in rather unusual and
extreme conditions. At the same time many questions regarding AM CVn
stars are unanswered. Below I list some of the outstanding questions,
their importance and sometimes their (partial) answers.

\subsection{Interpretation of V407 Vul, RXJ0806+15}

One of the main questions that people have tried to answer in the
recent years is what the nature of the two short-period candidates is.
They have been proposed to be either double degenerate polars
\citep{chm+98}, direct impact accretors \citep[][see
below]{ms02,rwc02}, double white dwarfs with an electrical interaction
like the Jupiter-Io system \citep{wcr02}, or face-on stream-fed
intermediate polars \citep{nhw02}. In the last two cases they would
not belong to the AM CVn stars. See for a discussion \citet{crw+04},
who conclude that the electrical star is the most viable model. Since
then the period derivatives have been confirmed and detection of
linear polarisation for RXJ0806+15 has been reported \citep{isr04}.

If the measured period derivative reflect the secular evolution of
these systems (which it tends not to do in interacting binaries) it
strongly argues against an AM CVn star interpretation.  The period
derivatives are roughly in agreement with what is expected for
detached double white dwarfs spiralling together due to GWR. However,
the inferred masses for V407 Vul \citep{str03} are too low for a
detached double white dwarf \citep[see Fig~1 in][]{npv+00}.
Furthermore, for a dipole magnetic field, the necessary geometry to
make the electrical star fit the observed X-ray light curve seems to
be very contrived (Barros et al., these proceedings).

For the intermediate polar model, the absence of emission lines, the
very soft X-ray spectrum and the absence of any cool component in the
optical and IR spectral shape of RXJ0806+15 \citep[limiting any donor
star to spectral types later than L6 or later][]{rbs04} are serious
problems. The apparent support for this model by the presence of a
G-star in the spectrum of V407 Vul is hampered by the fact that the
star does not show radial velocity variations more than
$\sim$10~km~s$^{-1}$, limiting the inclination to $<4^{\circ}$ for the
system parameters given in \citet{nhw02}, and even smaller if a more
realistic mass for the G9V star is taken.

In summary all proposed models seem to have problems explaining these
systems, and it is not clear yet if any of these models will be the
right one.

\subsection{Physics of the He accretion disc}

\begin{figure}[t]
\begin{center}
\psfig{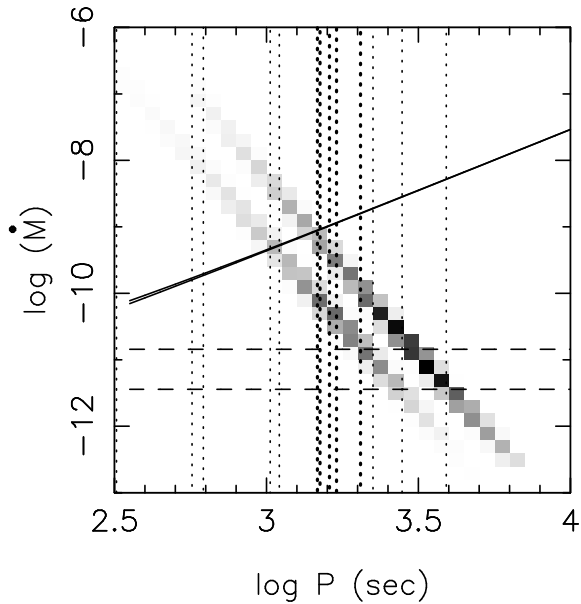}
\end{center}
\vspace*{-0.5cm}
\caption{Period mass transfer plot with the two tracks of the
  different formation scenarios and the stability criteria of
  \citet{to97} plotted. Above the solid lines the discs should be hot
  and stable, below the dashed lines (upper for an accretor mass of
  0.5 M$_\odot$, the lower for 1.0 M$_\odot$). The periods of the
  known systems are plotted as the dotted lines (thicker for outbursting
  systems). From \citet{npv+00}
  with updates. }
\label{fig:disc_stability}
\vspace*{-0.3cm}
\end{figure}

A unique property of AM CVn stars are their He dominated accretion
discs. Furthermore, the GWR driven mass transfer is expected to drop
by at least five orders of magnitude over the observed period range
(see Fig.~\ref{fig:P_Mdot}), allowing in principle the study of He
accretion discs as function of mass-transfer rate. \citet{to97}
studied the stability of He discs and concluded that for increasing
orbital period the AM CVn stars should go through a phase with a
stable hot discs (identified with the high-state systems), a phase
with an unstable disc (identified with the outbursting systems) and
settle down in a phase with a cool (neutral) stable disc (identified
with the low-state systems). The stability lines as given by
\citet{to97} actually match the then known observed properties well.
The new systems follow the trend and if SDSSJ1240-01 does indeed not
show outbursts, the transition to the stable cool discs happens
between 34 and 39 minutes. In Fig.~\ref{fig:disc_stability} I plot
the expected visible population of AM CVn stars, the stability criteria
and the periods of the observed systems \citep[after][]{npv+00}.

In recent years a number of groups have attempted to construct model
disc spectra in order to constrain the system parameters of the
observed objects. \citet{nss01} found relatively high mass transfer
rates of a few times $10^{-9}$~M$_\odot$~yr$^{-1}$ for AM CVn, HP Lib,
CR Boo and V803 Cen. \citet{ew00} modelled AM CVn and CR Boo and find
somewhat lower mass transfer rates
($\sim10^{-9}$~M$_\odot$~yr$^{-1}$). \citet{ndk04} modelled the
spectrum of AM CVn with the parameters of \citet{nss01} but do not
reproduce the relative strengths of the HeI 4026 and 4143 \AA\ lines.

Finally, there have been a number of recent SPH calculations reported
that are relevant for AM CVn stars \citep{wms00,ksh01,pea03}, coupling
their results to the observed (negative) superhumps. \citet{pea03}
suggests that AM CVn has a relatively high donor mass with a modest
magnetic field.

\subsection{Evolution and system parameters}

One of the promising ways to distinguish between the different
possible formation scenarios for the currently observed systems is to
study the chemical composition of the donor star \citep{nt02}. This
reveals itself through the chemical composition of the accretion disc
which dominates the light (except for the low-state systems, see
below). In case the donor is a white dwarf, its composition should be
that of a helium white dwarf or a carbon/oxygen white dwarf. The
descendants of the helium star channel should (at some point) show
produces of helium burning (i.e. carbon) in their spectrum, although
the theoretical models suggest most went though very little helium
burning \citet{npv+00}. Lastly, the systems formed from CVs with
evolved donors can show traces of H \citep{phr01}.

In none of the confirmed AM CVn systems is there any sign of H. The
spectral models suggest that only a tiny fraction of H ($10^{-5}$
by number) would be detectable in the spectrum. On the other hand, the
spectrum of GP Com shows a large overabundance of N \citep{mhr91}, as
is expected for a helium white dwarf donor or possibly a remnant from
an evolved CV donor if all the H is lost. CE315 shows a similar
effect. However, as is discussed above \citet{mhr91} suggest
additional N is accreted when its companion was an AGB star. The
low-state spectrum of CP Eri does show Si lines, suggesting it had
higher initial metallicity.

Recent XMM observations confirm the high N abundance and show Ne lines
in GP Com \citep{str04b}, while spectra of AM CVn, HP Lib and CR Boo
seem to show evidence for enhanced N, in case of AM CVn and HP Lib
again above the expected value of CNO processed material (Ramsay et
al., in prep). Finally, the newly discovered systems in which both
lines from the disc and lines from the accreting white dwarf are seen,
both show the Fe 5169 \AA\ line in the disc (and again no sign of H in
the absorption lines of the accreting white dwarf). 

It is generally assumed that the white dwarf donors obey the mass --
radius relation for zero-temperature white dwarfs
\citep[e.g.][]{npv+00} which is an extreme simplification.  Recent
calculations have become available of finite entropy low-mass white
dwarfs as donors in ultra-compact X-ray binaries \citep{db03} and AM
CVn stars (Deloye et al. in prep) will fill in the gap in our
knowledge here. Another very interesting recent development is the
application of the calculations of compressional heating of accreting
white dwarfs in CVs \citep{tb04} to the accreting white dwarfs in AM
CVn systems (see Bildsten, these proceedings). The result is that at
late times (long orbital periods) the accreting white dwarf will
dominate the light from the system. This is clearly the case in the
newly discovered systems showing the accreting white dwarf absorption
lines, but is also the case for GP Com and CE315. For the latter two,
this can be directly inferred from the effective temperature of the
continuum and the absolute magnitude (Bildsten et al., in prep).

\subsection{Population}


At present the extent of the Galactic population of AM CVn stars is
not (well) known. Theoretical calculations predict between $\sim10^7$
to more than $10^8$ in the Galaxy \citep[e.g.][]{ty96,npv+00,htp02}
but the uncertainties in both the efficiencies of the formation
scenarios and the selection effects are very large \citep{npv+00}.  In
any case the vast majority of the systems in the Galaxy should be at
the long period end, where the systems pile up at very low
mass-transfer rate.  Also, estimates based on the observed systems
\citep[e.g.][]{war95} are heavily affected by selection effects, both
with regard to the brightness of the objects as well as the fact that
almost all known sources have been found serendipitously. Only
recently, with the large surveys like the Sloan survey and the
dedicated AM CVn surveys will the latter point be remedied. The
increasing number of systems with distance determinations will greatly
help the question of the brightness of the systems as function of
orbital period.

We recently started looking in more detail at the selection effects
and calculated the expected number of detectable short-period AM CVn
systems, in particular to address the question of how likely it would
be to have two direct impact accretors in the Galaxy that are
currently visible.  We found that in order to expect a few of these
systems, we need to adopt one of the most optimistic models about the
total population in the Galaxy \citep{nyp03}. That would suggest there
should still be many more relatively bright longer period systems
undetected.

\section{AM CVn stars as GWR sources}

\begin{figure}[t]
\begin{center}
\psfig{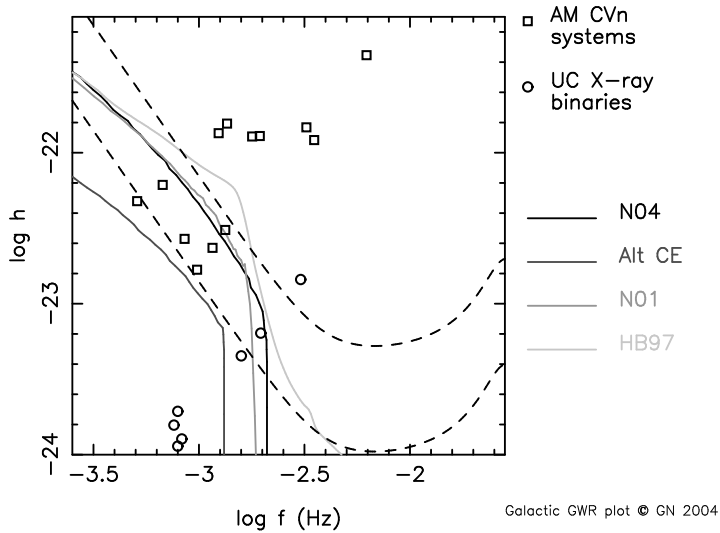}
\end{center}
\vspace*{-0.5cm}
\caption{Known AM CVn stars (open squares) and Ultra-compact X-ray
  binaries (open circles) in the GWR frequency - GWR strain amplitude
  plain. The dashed lines show the \textit{LISA} sensitivity for a one
  year mission giving a signal-to-noise ratio of 1 and 5 respectively.
  The Solid curves are different estimates of the average confusion
  limited double white dwarf background. Black: \citet{nyp03} (N04),
  dark grey: the same with the alternative common-envelope
  \citep{nvy+00} applied in all cases, lighter
  grey: \citet{nyp+00} (N01) and lightest grey: \citet{hb97} (HB97).  }
\label{fig:GalacticGWR}
\vspace*{-0.3cm}
\end{figure}

AM CVn stars are important sources of low-frequency GWR
\citep[e.g.][]{nel03}, a frequency range that will be probed by the
ESA/NASA \textit{LISA}
mission\footnote{http://sci.esa.int/science-e/www/area/index.cfm?fareaid=27\\http://lisa.jpl.nasa.gov/}.
They hardly contribute to the noise background \citep{hb00}, but they
are the only \textit{known} sources in this frequency band (although
many exotic sources like massive black hole mergers, compact objects
spiralling into massive black holes etc. are expected).
Fig.~\ref{fig:GalacticGWR} shows the known and candidate AM CVn stars
and the \textit{LISA} sensitivity. It is clear that even if the two
candidates are not AM CVn stars there are still seven objects that
will be easily detected (with the caveat that except for AM CVn the
distances are not well known).

A more detailed discussion of AM CVn stars as sources of GWR is given
in \citet{nyp03}, in particular the promising possibility to combine
GWR and electro-magnetic observations to study AM CVn stars. The
expected number of AM CVn stars that \textit{LISA} will detect
individually is many thousands \citep[although the data reduction
problem of how exactly all these signals are recovered from the data
is far from solved, e.g.][]{cl03,vw04}.

\section{Conclusions and outlook}

In this concluding section I summarise some of the questions that I
find interesting and briefly discuss (near) future developments.

\begin{description}
\item[Formation and properties of AM CVn stars]
The properties of the observed systems seem to be in rough agreement
with having a strongly decreasing mass transfer rate as function of
orbital period. The details of the mass transfer rate as orbital
period are currently studied in more detail, through more realistic
models of the donor stars. At the same time, predictions made by
calculating the effective temperature and brightness of the accreting
white dwarfs agree with the observations.  The question of which of
the three different formation channels led to the observed systems
will hopefully be answered in the relatively near future by detailed
studies of the chemical composition of the accreted material, both
theoretically as function of formation history (including possible
accretion in a previous mass transfer phase) and initial metallicity
and observationally by higher signal-to-noise ratio spectra. Current
preliminary conclusions are that there seems to be a quite large range
in initial metallicities (some quite low).  Secondly, there currently
is no evidence for H in any system, but there is a lot of evidence for
enhanced N, indicating CNO processing in the progenitors, i.e.  either
helium white dwarfs or evolved secondaries in CVs.

\item[AM CVn stars as astrophysical laboratories] The use of AM CVn
  stars as astrophysical laboratories is hampered by our lack of
  understanding both of the formation and population of systems in the
  Galaxy. Upcoming surveys (and hopefully theoretical developments
  that will allow us to compare the survey results directly with
  population studies) will probably greatly help sort out at least the
  population question. The fact that a number of distance estimates
  are available makes that questions like the brightness of the
  accretion discs and the accreting white dwarf (and thus their sizes)
  can be addressed. The disc instability model applied to helium discs
  has been mentioned above and the discovery of more systems will
  enable much more strict tests of this model than was previously
  possible. Finally, binary evolution constraints, in particular on
  the common envelope phase, which determines the mass ratios of the
  detached double white dwarfs \citep{nyp+00} which in turn determines
  the number of possible AM CVn progenitors. This is also relevant in
  the search for the progenitors of type Ia supernovae. Finally, the
  direct impact phase, which must exist, regardless of the
  identification of the two candidate AM CVn stars as direct impact
  accretors, will be an interesting aspect.
\item[AM CVn stars as GWR sources] AM CVn stars are currently the only
  known sources for the space based GWR detector \textit{LISA}. The
  expected number of individually detected sources is many thousands
  and the \textit{LISA} measurements will be mainly sensitive to the
  relatively rare short period systems that are difficult to access
  with classical observational techniques. 
\item[The nature of V407 Vul and RXJ0806+15] The nature of the two
  candidate short-period AM CVn systems is unclear. All proposed
  models seem to have problems explaining the observations. They are
  very interesting objects in any case and the observations scheduled
  in the near future will keep them in the spotlight. Hopefully either
  observational or theoretical advances will solve this puzzle soon.
  If not, \textit{LISA} will certainly do so.
\end{description}

\acknowledgments It is a pleasure to thank Tom Marsh, Paul Groot,
Danny Steeghs, Gijs Roelofs, Gavin Ramsay, GianLuca Israel, Lars
Bildsten, Chris Deloye, Dean Townsley, Patrick Woudt, Lev Yungelson,
Simon Portegies Zwart and Frank Verbunt for stimulating discussions
and for helping me to get an overview of the current status of AM CVn
stars.

\bibliographystyle{NBmn}
\bibliography{journals,binaries}

\begin{thebibliography}
\expandafter\ifx\csname natexlab\endcsname\relax\def\natexlab#1{#1}\fi

\bibitem[{{Bender} \& {Hils}(1997)}]{hb97}
{Bender} P.L., {Hils} D., 1997, Class. and Quantum Grav.,  14, 1439

\bibitem[{{Chornock} \& {Filippenko}(2003)}]{cf03}
{Chornock} R., {Filippenko} A.V., 2003, \iaucirc,  8084, 3

\bibitem[{{Cornish} \& {Larson}(2003)}]{cl03}
{Cornish} N.J., {Larson} S.L., 2003, \prd,  67, 103001

\bibitem[{Cropper et~al.(1998)Cropper, Harrop-Allin, Mason et~al.}]{chm+98}
Cropper M., Harrop-Allin M.K., Mason K.O., et~al., 1998, MNRAS,  293, L57

\bibitem[{Cropper et~al.(2004)Cropper, Ramsay, Wu \& Hakala}]{crw+04}
Cropper M., Ramsay G., Wu K., Hakala P., 2004, in Cropper M., Vrielman S.,
  eds., Proc Cape Town Workshop on magnetic CVs, ASP Conf. Series,
  astro-ph/0302240

\bibitem[{{Deloye} \& {Bildsten}(2003)}]{db03}
{Deloye} C.J., {Bildsten} L., 2003, \apj,  598, 1217

\bibitem[{El-Khoury \& Wickramasinghe(2000)}]{ew00}
El-Khoury W., Wickramasinghe D., 2000, A\&A,  358, 154

\bibitem[{{G{\" a}nsicke} et~al.(2003){G{\" a}nsicke}, {Szkody}, {de Martino}
  et~al.}]{gsd+03}
{G{\" a}nsicke} B.T., {Szkody} P., {de Martino} D., et~al., 2003, \apj,  594,
  443

\bibitem[{Groot et~al.(2001)Groot, Nelemans, Steeghs \& Marsh}]{gns+01}
Groot P.J., Nelemans G., Steeghs D., Marsh T.R., 2001, \apjl,  558, L123

\bibitem[{{Hakala} et~al.(2003){Hakala}, {Ramsay}, {Wu} et~al.}]{hrw+03}
{Hakala} P., {Ramsay} G., {Wu} K., et~al., 2003, \mnras,  343, 10

\bibitem[{Hils \& Bender(2000)}]{hb00}
Hils D., Bender P.L., 2000, ApJ,  537, 334

\bibitem[{{Hurley} et~al.(2002){Hurley}, {Tout} \& {Pols}}]{htp02}
{Hurley} J.R., {Tout} C.A., {Pols} O.R., 2002, \mnras,  329, 897

\bibitem[{Iben \& Tutukov(1991)}]{it91}
Iben I. Jr, Tutukov A.V., 1991, ApJ,  370, 615

\bibitem[{Israel et~al.(2004)}]{isr04}
Israel G., et~al., 2004, in Tovmassian G., Sion E., eds., Compact binaries in
  The Galaxy and beyond, volume~20 of \emph{RevMexAA (SC)}, p. 275

\bibitem[{{Israel} et~al.(1999){Israel}, {Panzera}, {Campana} et~al.}]{ipc+99}
{Israel} G.L., {Panzera} M.R., {Campana} S., et~al., 1999, \aap,  349, L1

\bibitem[{{Israel} et~al.(2002){Israel}, {Hummel}, {Covino} et~al.}]{ihc+02}
{Israel} G.L., {Hummel} W., {Covino} S., et~al., 2002, \aap,  386, L13

\bibitem[{{Israel} et~al.(2003){Israel}, {Covino}, {Stella} et~al.}]{ics+03}
{Israel} G.L., {Covino} S., {Stella} L., et~al., 2003, \apj,  598, 492

\bibitem[{Jha et~al.(1998)Jha, Garnavich, Challis, Kirshner \&
  Berlind}]{jgc+98}
Jha S., Garnavich P., Challis P., Kirshner R., Berlind P., 1998, \iaucirc,
  6983

\bibitem[{{Kato} et~al.(2000){Kato}, {Nogami}, {Baba}, {Hanson} \&
  {Poyner}}]{knb+00}
{Kato} T., {Nogami} D., {Baba} H., {Hanson} G., {Poyner} G., 2000, \mnras,
  315, 140

\bibitem[{{Kato} et~al.(2004){Kato}, {Stubbings}, {Monard} et~al.}]{ksm+04}
{Kato} T., {Stubbings} R., {Monard} B., et~al., 2004, \pasj,  56, 89

\bibitem[{{Kunze} et~al.(2001){Kunze}, {Speith} \& {Hessman}}]{ksh01}
{Kunze} S., {Speith} R., {Hessman} F.V., 2001, \mnras,  322, 499

\bibitem[{Kuulkers et~al.(2004)Kuulkers, Norton, Schwope \& Warner}]{kns+04}
Kuulkers E., Norton A., Schwope A., Warner B., 2004, in Lewin W., van~der Klis
  M., eds., Compact Stellar X-Ray Sources,, CUP

\bibitem[{Marsh et~al.(2004)Marsh, Nelemans \& Steeghs}]{mns02}
Marsh T., Nelemans G., Steeghs D., 2004, \mnras,  350, 113

\bibitem[{{Marsh} \& {Steeghs}(2002)}]{ms02}
{Marsh} T.R., {Steeghs} D., 2002, \mnras,  331, L7

\bibitem[{Marsh et~al.(1991)Marsh, Horne \& Rosen}]{mhr91}
Marsh T.R., Horne K., Rosen S., 1991, ApJ,  366, 535

\bibitem[{{Morales-Rueda} et~al.(2003){Morales-Rueda}, {Marsh}, {Steeghs}
  et~al.}]{mms+03}
{Morales-Rueda} L., {Marsh} T.R., {Steeghs} D., et~al., 2003, \aap,  405, 249

\bibitem[{Motch et~al.(1996)Motch, Haberl, Guillout et~al.}]{mhg+96}
Motch C., Haberl F., Guillout P., et~al., 1996, \aap,  307, 459

\bibitem[{{Nagel} et~al.(2003){Nagel}, {Dreizler} \& {Werner}}]{ndk04}
{Nagel} T., {Dreizler} S., {Werner} K., 2003, in de~Martino D., Kalytis R.,
  Silvotti R., Solheim J., eds., White Dwarfs, Proc. XIII Workshop on White
  Dwarfs, Kluwer, pp. 357--358

\bibitem[{{Nasser} et~al.(2001){Nasser}, {Solheim} \& {Semionoff}}]{nss01}
{Nasser} M.R., {Solheim} J.E., {Semionoff} D.A., 2001, \aap,  373, 222

\bibitem[{Nelemans(2003)}]{nel03}
Nelemans G., 2003, Class. Quantum Grav.,  20, S81

\bibitem[{{Nelemans} \& {Tout}(2003)}]{nt02}
{Nelemans} G., {Tout} C.A., 2003, in de~Martino D., Kalytis R., Silvotti R.,
  Solheim J., eds., White Dwarfs, Proc. XIII Workshop on White Dwarfs, Kluwer,
  pp. 359--360

\bibitem[{Nelemans et~al.(2000)Nelemans, Verbunt, Yungelson \&
  Portegies~Zwart}]{nvy+00}
Nelemans G., Verbunt F., Yungelson L.R., Portegies~Zwart S.F., 2000, A\&A,
  360, 1011

\bibitem[{Nelemans et~al.(2001{\natexlab{a}})Nelemans, Portegies~Zwart, Verbunt
  \& Yungelson}]{npv+00}
Nelemans G., Portegies~Zwart S.F., Verbunt F., Yungelson L.R.,
  2001{\natexlab{a}}, A\&A,  368, 939

\bibitem[{Nelemans et~al.(2001{\natexlab{b}})Nelemans, Steeghs \&
  Groot}]{nsp00}
Nelemans G., Steeghs D., Groot P.J., 2001{\natexlab{b}}, MNRAS,  326, 621

\bibitem[{Nelemans et~al.(2001{\natexlab{c}})Nelemans, Yungelson,
  Portegies~Zwart \& Verbunt}]{nyp+00}
Nelemans G., Yungelson L.R., Portegies~Zwart S.F., Verbunt F.,
  2001{\natexlab{c}}, A\&A,  365, 491

\bibitem[{Nelemans et~al.(2004)Nelemans, Yungelson \& Portegies~Zwart}]{nyp03}
Nelemans G., Yungelson L.R., Portegies~Zwart S.F., 2004, \mnras,  349, 181

\bibitem[{{Norton} et~al.(2004){Norton}, {Haswell} \& {Wynn}}]{nhw02}
{Norton} A.J., {Haswell} C.A., {Wynn} G.A., 2004, A\&A,  419, 1025

\bibitem[{Paczy\'nski(1967)}]{pac67}
Paczy\'nski B., 1967, Acta Astron.,  17, 287

\bibitem[{Patterson et~al.(1997)Patterson, Kemp, Shambrook et~al.}]{pks+97}
Patterson J., Kemp J., Shambrook A., et~al., 1997, PASP,  109, 1100

\bibitem[{{Patterson} et~al.(2000){Patterson}, {Walker}, {Kemp}
  et~al.}]{pwk+00}
{Patterson} J., {Walker} S., {Kemp} J., et~al., 2000, \pasp,  112, 625

\bibitem[{{Patterson} et~al.(2002){Patterson}, {Fried}, {Rea} et~al.}]{pfr+02}
{Patterson} J., {Fried} R.E., {Rea} R., et~al., 2002, \pasp,  114, 65

\bibitem[{{Pearson}(2003)}]{pea03}
{Pearson} K.J., 2003, \mnras,  346, L21

\bibitem[{{Podsiadlowski} et~al.(2003){Podsiadlowski}, {Han} \&
  {Rappaport}}]{phr01}
{Podsiadlowski} P., {Han} Z., {Rappaport} S., 2003, MNRAS,  340, 1214

\bibitem[{Ramsay et~al.(2000)Ramsay, Cropper, Wu, Mason \& Hakala}]{rcw+00}
Ramsay G., Cropper M., Wu K., Mason K.O., Hakala P., 2000, MNRAS,  311, 75

\bibitem[{{Ramsay} et~al.(2002{\natexlab{a}}){Ramsay}, {Hakala} \&
  {Cropper}}]{rhc02}
{Ramsay} G., {Hakala} P., {Cropper} M., 2002{\natexlab{a}}, \mnras,  332, L7

\bibitem[{{Ramsay} et~al.(2002{\natexlab{b}}){Ramsay}, {Wu}, {Cropper}
  et~al.}]{rwc02}
{Ramsay} G., {Wu} K., {Cropper} M., et~al., 2002{\natexlab{b}}, \mnras,  333,
  575

\bibitem[{Reinsch et~al.(2004)Reinsch, Burwitz \& Schwarz}]{rbs04}
Reinsch K., Burwitz V., Schwarz R., 2004, in Tovmassian G., Sion E., eds.,
  Compact binaries in The Galaxy and beyond, volume~20 of \emph{RevMexAA (SC)},
  p. 122

\bibitem[{{Roelofs} et~al.(2004){Roelofs}, {Groot}, {Steeghs} \&
  {Nelemans}}]{rgs+04}
{Roelofs} G., {Groot} P., {Steeghs} D., {Nelemans} G., 2004, in Tovmassian G.,
  Sion E., eds., Compact binaries in The Galaxy and beyond, volume~20 of
  \emph{RevMexAA (SC)}, p. 254

\bibitem[{{Ruiz} et~al.(2001){Ruiz}, {Rojo}, {Garay} \& {Maza}}]{rrg+01}
{Ruiz} M.T., {Rojo} P.M., {Garay} G., {Maza} J., 2001, \apj,  552, 679

\bibitem[{{Seetha} et~al.(2000){Seetha}, {Ashoka}, {Raj} \&
  {Kasturirangan}}]{sar+00}
{Seetha} S., {Ashoka} B.N., {Raj} E., {Kasturirangan} K., 2000, Bulletin of the
  Astronomical Society of India,  28, 247

\bibitem[{Skillman et~al.(1999)Skillman, Patterson, Kemp et~al.}]{spk+99}
Skillman D.R., Patterson J., Kemp J., et~al., 1999, PASP,  111, 1281

\bibitem[{Smak(1967)}]{sma67}
Smak J., 1967, Acta Astron.,  17, 255

\bibitem[{Solheim(1995)}]{sol95}
Solheim J.E., 1995, Baltic Astronomy,  4, 363

\bibitem[{{Solheim}(2003)}]{sol03}
{Solheim} J.E., 2003, in {de Martino} D., {Silvotti} R., {Solheim} J.E.,
  {Kalytis} R., eds., White Dwarfs, volume 105 of \emph{NATO Science Series II
  -- Mathematics, Physics and Chemistry}, Kluwer, p. 299

\bibitem[{{Strohmayer}(2002)}]{str03}
{Strohmayer} T.E., 2002, \apj,  581, 577

\bibitem[{{Strohmayer}(2003)}]{str03b}
{Strohmayer} T.E., 2003, \apjl,  593, L39

\bibitem[{{Strohmayer}(2004{\natexlab{a}})}]{str04c}
{Strohmayer} T.E., 2004{\natexlab{a}}, \apj,  in press, astro-ph/0405203

\bibitem[{{Strohmayer}(2004{\natexlab{b}})}]{str04a}
{Strohmayer} T.E., 2004{\natexlab{b}}, \apj,  610, 416

\bibitem[{{Strohmayer}(2004{\natexlab{c}})}]{str04b}
{Strohmayer} T.E., 2004{\natexlab{c}}, \apjl,  608, L53

\bibitem[{{Thorstensen}(2003)}]{tho03}
{Thorstensen} J.R., 2003, AJ,  126, 3017

\bibitem[{{Townsley} \& {Bildsten}(2004)}]{tb04}
{Townsley} D.M., {Bildsten} L., 2004, \apj,  600, 390

\bibitem[{Tsugawa \& Osaki(1997)}]{to97}
Tsugawa M., Osaki Y., 1997, PASJ,  49, 75

\bibitem[{Tutukov \& Yungelson(1996)}]{ty96}
Tutukov A.V., Yungelson L.R., 1996, MNRAS,  280, 1035

\bibitem[{{Vecchio} \& {Wickham}(2004)}]{vw04}
{Vecchio} A., {Wickham} E.D.L., 2004, Class. and Quantum Grav.,  21, 661

\bibitem[{Warner(1995)}]{war95}
Warner B., 1995, Ap\&SS,  225, 249

\bibitem[{{Warner}(2004)}]{war04}
{Warner} B., 2004, \pasp,  pp. 115--132

\bibitem[{{Warner} \& {Woudt}(2002)}]{ww02}
{Warner} B., {Woudt} P.A., 2002, \pasp,  114, 129

\bibitem[{Webbink(1984)}]{web84}
Webbink R.F., 1984, ApJ,  277, 355

\bibitem[{Whitehurst(1988)}]{whi88}
Whitehurst R., 1988, MNRAS,  232, 35

\bibitem[{{Wood} et~al.(2000){Wood}, {Montgomery} \& {Simpson}}]{wms00}
{Wood} M.A., {Montgomery} M.M., {Simpson} J.C., 2000, \apjl,  535, L39

\bibitem[{{Wood} et~al.(2002){Wood}, {Casey}, {Garnavich} \& {Haag}}]{wcg+02}
{Wood} M.A., {Casey} M.J., {Garnavich} P.M., {Haag} B., 2002, \mnras,  334, 87

\bibitem[{{Woudt} \& {Warner}(2003{\natexlab{a}})}]{ww03}
{Woudt} P.A., {Warner} B., 2003{\natexlab{a}}, in {White Dwarfs: Galactic and
  Cosmological Probes, proceedings of IAU JD5}, astro-ph/0310494

\bibitem[{{Woudt} \& {Warner}(2003{\natexlab{b}})}]{ww03a}
{Woudt} P.A., {Warner} B., 2003{\natexlab{b}}, \mnras,  345, 1266

\bibitem[{{Wu} et~al.(2002){Wu}, {Cropper}, {Ramsay} \& {Sekiguchi}}]{wcr02}
{Wu} K., {Cropper} M., {Ramsay} G., {Sekiguchi} K., 2002, \mnras,  331, 221

\end{thebibliography}

\end{document}